\documentclass[runningheads]{llncs}
\usepackage[T1]{fontenc}
\usepackage{graphicx}

\usepackage{graphicx}
\usepackage{longtable}
\usepackage{amsfonts}
\usepackage{amssymb}
\usepackage{bm}
\usepackage[utf8]{inputenc}
\usepackage[english]{babel}
\usepackage{xstring}
\usepackage{setspace}
\usepackage{array}
\usepackage{subfig}
\usepackage{enumitem}
\usepackage{multirow}
\usepackage{algorithmic}
\usepackage{dutchcal} 			% for \VF formulas (this package has conflict with calligraphy font. 17-11-2020)
\usepackage{ctable} 				% for \specialrule command
\usepackage{comment}
\usepackage[ruled,vlined,linesnumbered]{algorithm2e}
\usepackage{amsmath,amsfonts,amssymb}
\usepackage{calligra}
\usepackage{balance}
\usepackage{blkarray}
\newcommand{\V}[1] {\textsf{#1}} 			% Values in SIS program
\newcommand{\Bold}[1] {\bm{#1}} 			% Bold font for vectors

\newcommand{\Sensor}[1]{
    \IfEqCase{#1}{
       {IW512}{L1}
       {IW576}{L2}
       {IW544}{L3}
       {IW554}{P1}
       {IW560}{P2}
        {#1}{#1}
    }
}

\newcommand{\TE}[1]{{\scriptsize\sffamily #1}} %Table Entry Font and size
\newcolumntype{M}[1]{>{\centering\arraybackslash}m{#1}}
\newcolumntype{N}{@{}m{0pt}@{}}
\newcommand{\Eq}[1]{(\ref{#1})}

\newcommand{\Fig}[1]{Fig. \ref{#1}}
\newcommand{\Table}[1]{Table \ref{#1}}
\newcommand{\Sec}[1]{Section \ref{#1}}

\begin{document}

\raggedbottom
\title{SIL Allocation for Mitigation Safety Functions} 
\author{Hamid Jahanian}
\institute{FS Expert (TÜV Rheinland) \#266/16-SIS\\UGL, Sydney, Australia\\ 
\email{hamid.jahanian@ugllimited.com}}

\maketitle

\begin{abstract}
SIL (Safety Integrity Level) allocation plays a pivotal role in evaluating the significance of Safety Functions (SFs) within high-risk industries. The outcomes of a SIL allocation study determine the design specifications necessary to uphold the Probability of Failure on Demand (PFD) below permissible limits, thus managing risk effectively. While extensive research has focused on SIL allocation for preventive SFs, there is a noticeable gap in attention towards mitigation SFs. To address this gap, this paper discusses the shortcomings of current methods and proposes a new approach to overcome them. The principles of the proposed method are substantiated by detailed mathematical formulation and the practical application of the method is demonstrated through a case study in a road tunnel project.

\end{abstract}

\section{Introduction}\label{Sec_Intro}

The Safety Integrity Level (SIL) is a discrete number ranging from 1 to 4, representing the importance level of Safety Functions (SFs) \cite{Ref_183}. A higher SIL indicates a greater significance of the SF in preventing hazardous events or mitigating their consequences. The process of assigning SIL ratings to SFs involves an analysis that evaluates potential hazards, estimates their probability of occurrence, and assesses the severity of their consequences. The outcome of this analysis often sets a target Probability of Failure on Demand (PFD), based on which a target SIL is determined for the corresponding SF, as defined by the standard \cite{Ref_186}.

Assigning appropriate SILs to SFs is crucial. A higher SIL typically requires more sophisticated hardware with a lower failure rate, as well as stricter engineering, operational, and maintenance practices specified in functional safety standards, particularly IEC 61508 \cite{Ref_183}. Failure to conduct an effective analysis to establish the appropriate SIL can lead to unnecessary costs if the risk is overestimated or compromise safety if it is underestimated.

\begin{figure}[!h]  
\begin{center}
\includegraphics[scale=0.5]{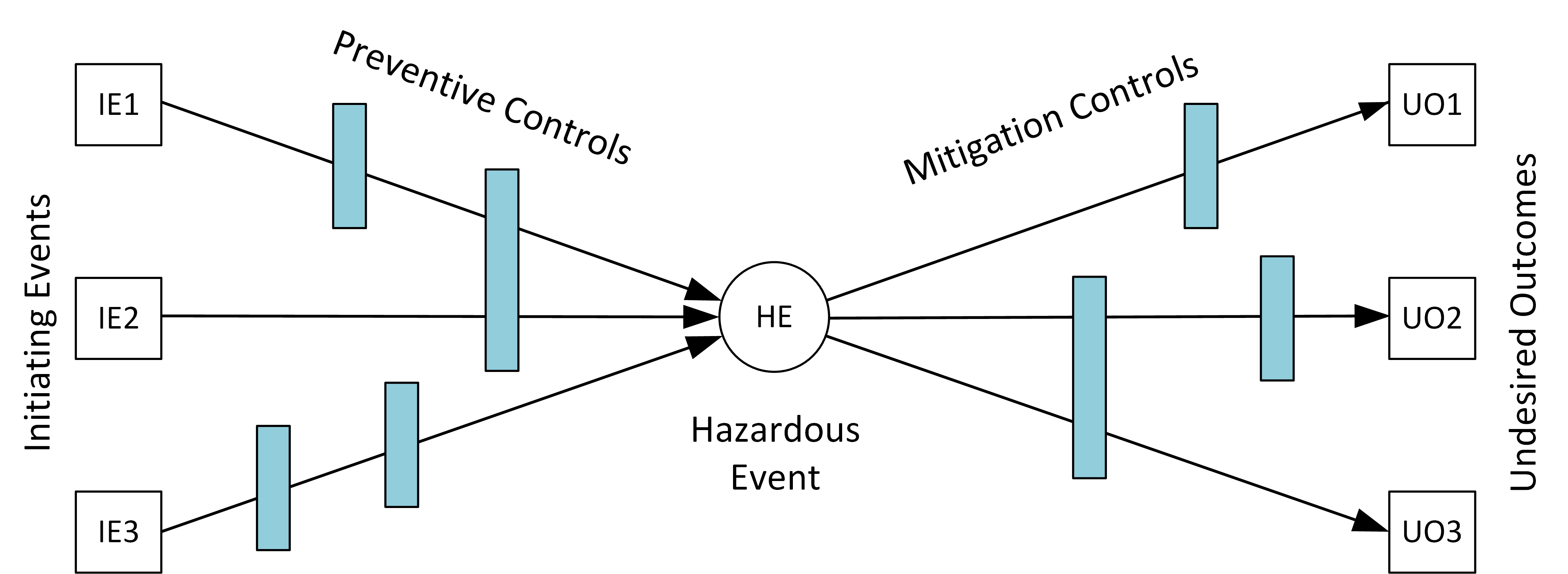}
\end{center}
\caption{Bowtie Accident Model}
\label{Fig_Bowtie}
\end{figure}

Safety Functions can be classified into two primary categories: preventive functions and mitigation functions. Preventive functions are designed to operate before a Hazardous Event (HE) occurs, aiming to prevent it from happening. Conversely, mitigation functions come into action after a hazardous event has occurred, focusing on minimizing the consequences. The relationship between these functions and the HE is often depicted in the form of a Bowtie diagram, with the HE in the centre, the initiating events and preventive controls on the left, and mitigation controls and undesired outcomes on the right; see \Fig{Fig_Bowtie}. 

While SIL allocation for preventive SFs has been extensively studied and addressed, there has been surprisingly limited focus on allocating SIL for mitigation SFs. To bridge this gap, this paper aims to review the methodological challenges and propose remedies. We first argue how the differences between mitigation SFs and preventive SFs render existing methods inefficient for allocating SIL to mitigation functions. We then present a new method that offers a more relevant and accurate estimation of the target PFD while addressing challenges such as multiple consequence categories and independence between mitigation subsystems. Our method examines all possible states of the system at the time of the event and incorporates all potential consequences that may arise from the hazardous event.

This paper makes contributions to the field of safety engineering. It addresses critical challenges in SIL determination for mitigation SFs and introduces a new method. The method is grounded in detail mathematical formulations and demonstrated through a realistic case study. By addressing this research gap, this paper aims to advance safety engineering practices.

The structure of this paper is as follows: Section \ref{Sec_Background} provides an overview of current practices of SIL allocation. In Section \ref{Sec_PFDFormul}, we formulate our new SIL allocation method. To demonstrate the practical application of this method, we present a case study from the road tunnel industry in Section \ref{Sec_CaseStudy}. With the new method presented, Section \ref{Sec_LitReview} provides a comparative review with the existing literature, and Section \ref{Sec_Discussion} revisits the initial question and explains how the new approach addresses the shortcomings of the current methods. This section also discusses additional aspects of SIL determination in relation to our method. Finally, in Section \ref{Sec_Conclusion}, we summarise our contributions and suggest future research directions to build upon the concepts and methods presented.

The following notations are used in this paper:
\begin{align}
l &: \textit{ number of subsystems} \nonumber \\
m &: \textit{ number of system functions} \nonumber \\
n &: \textit{ number of consequence segments} \nonumber \\
\Bold{\Psi}=[\psi_{ij}]_{2^l \times l} &: \textit{ possible system states} \nonumber \\
\Bold{\Phi}=[\phi_{ik}]_{2^l \times m} &: \textit{ possible function states} \nonumber \\
\Bold{\Gamma}=[\gamma_{ih}]_{2^l \times n} &: \textit{ mapping system states to consequences} \nonumber \\
\gamma_{h} : (i, \Bold{\Phi}) \mapsto \gamma_{ih} &: \textit{ a logic function mapping $\Bold{\Phi}_{i}$ to $\gamma_{ih}$}  \nonumber \\
\Bold{\omega}=[\omega_{i}]_{2^l \times 1} &: \textit{ system state frequencies} \nonumber \\
\Bold{F}=[f_{jk}]_{l \times m} &: \textit{ mapping subsystems to functions} \nonumber \\
\Bold{p}=[p_{j}]_{1 \times l} &: \textit{ PFD of subsystems} \nonumber \\
\Bold{w}=[w_{h}]_{1 \times n} &: \textit{ estimated consequence frequencies} \nonumber \\
\Bold{\overline{w}}=[\overline{w}_{h}]_{1 \times n} &: \textit{ tolerable consequence frequencies} \nonumber \\
\Bold{c}=[c_{h}]_{1 \times n} &: \textit{ consequence severity values} \nonumber \\
r &: \textit{ collective estimated level of risk} \nonumber \\
\overline{r} &: \textit{ collective tolerable level of risk} \nonumber  \\
w_{IE} &: \textit{ frequency of the initiating event} \nonumber \\
w_{HE} &: \textit{ frequency of the hazardous event} \nonumber \\
c_{HE} &: \textit{ hazardous event consequence} \nonumber 
\end{align}

\section{SIL Allocation}\label{Sec_Background}

SIL allocation can be approached through qualitative or quantitative methods. While qualitative methods directly determine target SIL ratings, quantitative methods utilise an intermediate measure such as PFD, Risk Reduction Factor (RRF), or frequency of failure per hour (PFH). In general, both PFD\footnote{RRF is simply the inverse of PFD: $RRF=1/PFD$.} and PFH represent \emph{dangerous} failures, where the failure of SFs can result in intolerable safety consequences. The relationship between these continuous measures and the discrete scale of SIL is defined by the IEC 61508 standard (see Tables 2 and 3 in \cite{Ref_183}).

One commonly practised qualitative method is the Risk Graph (RG) \cite{Ref_187,Ref_359}. This method involves selecting and setting several parameters contributing to the level of risk. These parameters typically include the consequence of a hazardous event (C), the frequency and exposure time to the hazard (F), the possibility of failing to avoid the hazard (P), and the probability of the unwanted occurrence (W). By considering and evaluating these parameters, an appropriate SIL can be determined based on a preset risk chart.

As an example of quantitative methods, Layers of Protection Analysis (LOPA) \cite{Ref_352,Ref_187} is extensively employed for SIL allocation in the process industry. Primarily designed for generic analysis of protection layers, LOPA uses a straightforward approach to calculate the frequency of a hazardous event ($w_{HE}$) based on the frequency of the initiating event ($w_{IE}$) and the probability of failure of the protection layers ($p_i$s). In a simplified form, with only one possible initiating event, this calculation can be expressed as follows:
\begin{align}
&w_{HE}=w_{IE} \prod_i p_i \label{Eq_LOPASimp}
\end{align}

Once the frequency of hazardous event is estimated, it can be compared to a tolerable frequency of hazardous event to identify the gap and determine if further control measures are needed to reduce the risk. Where the control measure is a Safety Function, the gap also reflects the target PFD.

LOPA is commonly known as a \emph{semi}-quantitative analysis method. It incorporates a set of rules for the protection layers, which, although not always easy to satisfy, greatly simplify the calculation process. Specifically, LOPA necessitates the independence of the protection layers both from each other and from the initiating events. This condition is clearly needed for utilizing the product relation \Eq{Eq_LOPASimp}.

In scenarios where dependencies between protection layers cannot be disregarded, alternative methods like Fault Tree Analysis (FTA) \cite{Ref_182,Ref_168} should be utilised to calculate $w_{HE}$. The estimated frequency can then be compared to the tolerable frequency, just as it would be done when using LOPA.

FTA requires the use of specialised analysis tools and expertise in their application. Nevertheless, this quantitative method is capable of calculating complex interrelations between initiating events and protection layers, providing a more comprehensive assessment of the system.

%Risk Graph is a qualitative method and does not delve into the role of protection measures at all. LOPA and FTA are scenario-based and do assess the role of individual protection mechanisms; however, this assessment is limited to the preventive measures, and it does not address the role of mitigation barriers between the hazardous event and its ultimate consequences. 

Event Tree Analysis (ETA) \cite{Ref_204,Ref_353} is a method used for post-event analyses. Unlike FTA, which begins with multiple ``basic events'' to determine the likelihood of a single ``top event,'' the ETA method starts with one initiating event - the hazardous event - and assesses the probabilities of multiple outcomes considering the success and failure of all post-event protection layers. The frequency of each outcome is determined by multiplying the probabilities associated with the branches connecting the initiating event to the specific outcome, and the frequency of the initiating event itself. This is similar to the LOPA calculations shown earlier in Equation \Eq{Eq_LOPASimp}. 

Mitigation SFs are post-event in nature, making ETA a valuable method for their assessment. However, ETA alone is insufficient for SIL allocation, as additional steps are necessary to gauge the overall risk based on estimated outcomes - a task that can be challenging in large-scale analyses. More importantly, ETA, too, depends on a fundamental condition: the independence of protection layers, which is often infeasible in complex, interdependent systems.

The method introduced in this paper aims to overcome these limitations, providing a solution that can be applied across various scenarios, regardless of analysis scale or subsystem dependencies.

\section{A New Method}\label{Sec_PFDFormul}

\textbf{Risk} is often defined as the combination of likelihood and severity of consequences (see \cite{Ref_186}, for instance). Likelihood may be expressed in the form of probability or frequency of occurrence, whereas severity can take any scale that expresses the extent of damage (e.g., monetary values). Where both the likelihood and severity are expressed in numerical terms, one common way to \emph{combine} the two is to multiply them as follows:

\begin{align}
&r = w \cdot c \label{Eq_RiskGeneral} 
\end{align}

Here, $w$ represents the frequency of occurrence, $c$ the severity of consequence, and $r$ the level of risk. The estimated risk can then be compared to a tolerable risk threshold $\overline{r}$, which is defined based on the given consequence severity. The goal then is to ensure that the estimated risk is below the tolerable level:
\begin{align}
&r(w, c) \leq \overline{r}(c) \label{Eq_RiskMeasGeneral}
\end{align}

When examining a preventive SF, the right-hand side of the Bowtie model (\Fig{Fig_Bowtie}) can be condensed into one overall $c_{HE}$ because preventive SFs have no impact on the right-hand side. Conversely, when the SF being studied serves a mitigation function, we can condense the left-hand side of the Bowtie into an estimated $w_{HE}$ and focus on the SF's influence on the ultimate consequence. 

Studying mitigation functions requires measuring the likelihood and consequences of multiple possible outcomes. To make the evaluation easier, the range of consequences is often divided into smaller \textbf{segments}. For instance, if the consequence severity ranges between 0 and 10, the range may be divided into 5 segments, each with a length of 2 units. 

Let $n$ be the number of all consequence segments and $c_h$, $w_h$, and $\overline{w}_h$ the consequence severity, the estimated frequency of occurrence, and the tolerable frequency of occurrence for segment $h$, respectively. The risk can then be evaluated in one of the following two ways, depending on the suitability of the approach and the decision of those involved in the risk evaluation process:
\begin{itemize}
\item Either $w_h$ to be below its corresponding $\overline{w}_h$ for all $h$s:
\begin{align}
&w_h \leq \overline{w}_h, ~~ for ~ 1 \leq h \leq n \label{Eq_RiskDiscA}
\end{align}
\item or the collective estimated risk to be below the collective tolerance threshold:
\begin{align}
&r=\sum_{h=1}^{n} w_h  c_h, ~~ \overline{r}=\sum_{h=1}^{n} \overline{w}_h  c_h, ~~ r \leq \overline{r} \label{Eq_RiskDiscB}
\end{align}
\end{itemize}

In some analyses, numerical expression of consequence is replaced by word labels like ``Minor'' and ``Major.'' Using word labels eliminates the unpleasant need for putting price tags on human consequences like loss of life, and it also aligns with the subjective nature of risk estimation, which may not be as obvious when using objective numbers. In such studies, one of the following two approaches may be taken to evaluate the risk:
\begin{itemize}
\item Either $w_h$s are estimated and compared with individual $\overline{w}_h$s, similar to \Eq{Eq_RiskDiscA},
\item or ``weight'' values are assigned to $c_h$s, as a replacement for the actual consequence severity values, and then a collective risk measure is evaluated as we did in \Eq{Eq_RiskDiscB}.
\end{itemize}

We use the term \textbf{mitigation function} to refer to a function that helps reduce risk only by decreasing the severity of consequences, without affecting the probability of the accident \cite{Ref_368}. A function that reduces the probability of an accident is considered a \textbf{preventive function}. We also use the term \textbf{mitigation system} to refer to the system that performs a specific set of mitigation functions. 

As an example, consider a fire incident in a road tunnel environment. The fire itself can result in casualties and damage to asset, whilst its smoke poses high risk on those who are confined in the tunnel even when distant from the fire location. The typical mitigation functions in this example are fire suppression, smoke extraction, and the evacuation of road users; and the mitigation system is the overall tunnel operation system, which enables the execution of these functions. 

A mitigation system comprises a set of \textbf{subsystems}. Each function relies on a number of these subsystems, some of which may be shared among multiple functions. Subsystems are assumed to be independent of each other.\footnote{The subsystem independence condition enables us to apply the product formula for calculating the probability of system states, and it is a crucial factor in defining system breakdown.} Let us consider a mitigation system composed of $l$ independent subsystems used to perform $m$ mitigation functions. To define the relationships between subsystems and functions, we introduce the \textbf{mapping matrix} $\Bold{F}$, as follows:
\begin{align}
&\Bold{F}=\bigl[f_{jk}]_{l \times m}, ~~ f_{jk} \in \{0, 1\} \label{Eq_MapF}
\end{align}
where $f_{jk}=1$ indicates that the $j$th subsystem is one of the subsystems that need to be functional for the $k$th function to succeed; and $f_{jk}=0$ indicates that the $j$th subsystem is not a part of the $k$th function.\footnote{Some safety standards present a simpler concept of Safety Functions and Safety-Related Systems and Subsystems (e.g., IEC 61511 \cite{Ref_190}). It should be noted, however, that these standards focus on preventive functions. When multiple mitigation functions are deployed that utilise shared subsystems, the relationship between functions and subsystems should be formulated using means such as the mapping matrix $\mathbf{F}$.}

A subsystem may be in either \textbf{available} or \textbf{unavailable} state at any given time, where available means that the subsystem is all functional and can perform as expected if there is a demand, and unavailable means otherwise. Given this binary state of the subsystems, a system with $l$ subsystems can only be in one of $2^l$ possible states at any given time. We use the \textbf{system state} matrix $\Bold{\Psi}$ to show all possible states of the system with respect to the availability of its subsystems:
\begin{align}
&\Bold{\Psi}=[\psi_{ij}]_{2^l \times l}, ~~ \Bold{\Psi}_{i}=[\psi_{ij}]_{1 \times l}, ~~ \psi_{ij} \in \{0, 1\}
\end{align}
where $\psi_{ij}=1$ means that the subsystem $j$ is considered available when the system is in state $i$, and $\psi_{ij}=0$ means that the subsystem $j$ is unavailable in state $i$. Here, $\psi_{ij}$ represents the state of one subsystem, $\Bold{\Psi}_{i}$ one possible state of the system, and $\Bold{\Psi}$ the set of all possible system states.

The availability state of the subsystems determines the success/failure state of the system functions. A function is considered \textbf{successful} if all the subsystems that are required to perform the function are available. Corresponding to the $2^l$ system states, we use the \textbf{function state} matrix $\Bold{\Phi}$ to show the success/failure states of the functions in various system states:
\begin{align}
&\Bold{\Phi}=[\phi_{ik}]_{2^l \times m}, ~~ \Bold{\Phi}_{i}=[\phi_{ik}]_{1 \times m}, ~~ \phi_{ik} \in \{0, 1\}
\end{align}
where $\phi_{ik}=1$ means that the $k$th function is successful when the system is in state $i$, and $\phi_{ik}=0$ means that the $k$th function is in failure state when the system in state $i$. The matrix $\Bold{\Phi}$ can be calculated based on $\Bold{\Psi}$ and $\Bold{F}$:
\begin{align}
&\Bold{\Phi}=\Bold{\Psi} \textcircled{$\wedge$} \Bold{F} \label{eq_MatrixAND}
\end{align}
with operator $\textcircled{$\wedge$}$ representing a special conjunction between $\Bold{\Psi}$ and $\Bold{F}$ as defined below:
\begin{align}
&\phi_{ik}= \bigwedge_{j=1}^{l} (\psi_{ij} \vee \neg f_{jk}) \label{eq_MatrixAND1}
\end{align}
This conjunction simply tests if all the subsystems that are required to perform the $k$th function are available when the system is in state $i$.\footnote{We use the 0/1 convention for Boolean logic instead of false/true. Thus the same real-valued 0 and 1 in  $\Bold{\Psi}$ and $\Bold{F}$ are combined in logical terms too.} 

The ultimate consequence of an HE is determined by the success/failure of the mitigation functions, which are in turn dependent on the availability/unavailability of subsystem. We introduce the \textbf{state consequence} matrix $\Bold{\Gamma}$ to express the consequence severity if the HE occurs when the system is in a given state:
\begin{align}
&\Bold{\Gamma}=[\gamma_{ih}]_{2^l \times n}, ~~ \Bold{\Gamma}_{i}=[\gamma_{ih}]_{1 \times n}, ~~ \gamma_{ih} \in \{0, 1\}
\end{align}
where $\gamma_{ih}=1$ indicates that when the system is in state $i$, the severity of consequence will be in segment $h$. 

The total number of states may be large. In the case study we present in \Sec{Sec_CaseStudy}, there are 10 subsystems, which can put the system in 1024 different states. Identifying the right consequence for each individual state may be tedious, unrepeatable, and prone to errors. In our method, this problem is solved by using logic functions to derive $\Bold{\Gamma_{i}}$ from the function state $\Bold{\Phi}_i$: 
\begin{align}
&\gamma_{ih}=\gamma_{h}(i, \Bold{\Phi})
\end{align}

Thus, the value of $\gamma_{ih}$ in matrix $\Bold{\Gamma}$ is determined by the logic function $\gamma_{h}()$ over the $i$th row in matrix $\Bold{\Phi}$. These functions are decided and defined by the analyst, and based on the application context. We demonstrate this as part of the case study in \Sec{Sec_CaseStudy}.

The second element of risk is likelihood. In our model, this is estimated based on the probability of system being in state $i$, which can in turn be calculated based on the probability of individual subsystems being available or unavailable. 

Let $p_j$ represent the probability of subsystem $j$ being unavailable. We define the PFD vector $\Bold{p}$ as follows:
\begin{align}
&\Bold{p}=[p_{j}]_{1\times l}, ~~ 0 \leq p_{j} \leq 1 \label{Eq_PFD}
\end{align}
 
Given that the $l$ subsystems are independent, the probability that the system is in state $i$ can be calculated as follows:
\begin{align}
&\prod_{j=1}^{l} \bigl(\psi_{ij} (1-p_j) + (1-\psi_{ij}) p_j\bigr) \label{Eq_OneFreq}
\end{align}
Let $\omega_i$ be the frequency of state $i$ and $\Bold{\omega}$ the \textbf{state frequency} vector, containing the frequency of occurrence of all system states. Hence:
\begin{align}
\Bold{\omega}&=[\omega_{i}]_{2^l \times 1}, ~~ 0 \leq \omega_{i}  
\end{align}
Using \Eq{Eq_OneFreq}, the frequency of the HE occurring when the system is in state $i$ will be calculated as follows:
\begin{align}
&\omega_i =  w_{HE} \prod_{j=1}^{l} \bigl(\psi_{ij} (1-p_j) + (1-\psi_{ij}) p_j\bigr) \label{Eq_SecenFreq}
\end{align}
Since the system can be in one state at any given time, we also observe that:
\begin{align}
&\sum_{i=1}^{2^l} \omega_i = w_{HE}
\end{align}

The values of $\omega_i$s represent the likelihood of the HE occurring in individual system states. To estimate the risk, we need to calculate the likelihood of the consequence segments. Let $w_h$ represent the frequency of HEs that result in consequence segment $h$ and $\Bold{w}$ the \textbf{consequence segment frequency} vector, as follows:
\begin{align}
\Bold{w}&=[w_{h}]_{1 \times n},  ~~ 0 \leq w_{h}
\end{align}
Using state consequence $\Bold{\Gamma}$ and  state frequency $\Bold{\omega}$, we can calculate the consequence frequency $\Bold{w}$ as follows:
\begin{align}
\Bold{w}&= \Bold{\omega}^T \times \Bold{\Gamma}
\end{align}

Recall that we could choose to compare the risk either in individual segments or as a collective figure. When using \Eq{Eq_RiskDiscA}, we need to compare the individual estimated consequence frequencies with their corresponding tolerance thresholds, which we express here by \textbf{segment tolerance frequency} vector $\Bold{\overline{w}}$ as defined below:
\begin{align}
\Bold{\overline{w}}&=[\overline{w}_{h}]_{1 \times n},  ~~ 0 < \overline{w}_{h}
\end{align}

As set earlier by \Eq{Eq_RiskDiscA}, the following condition should hold for the estimated risk to be considered tolerable:
\begin{align} 
&w_h \leq \overline{w}_h, ~~ for ~ 1 \leq h \leq n  \label{Eq_RiskTolInd}
\end{align}

When using a collective risk comparison, we need to assign values to consequence segments and combines them with their corresponding frequencies. As said earlier, the consequence values may be expressed by the scale of loss (e.g., monetary figures or casualty numbers) or in the form of ``weights.'' The latter provides a relative comparison between the segments while the former gives an estimation of the actual loss. Nonetheless, we use $\Bold{c}$ to express the value of \textbf{consequence severity}, as defined below:
\begin{align}
\Bold{c}&=[c_{h}]_{1 \times n},  ~~ 0<c_{min} \leq c_h \leq c_{max} 
\end{align}
Here, $c_{min}$ and $c_{max}$ show some realistic boundaries for the consequences that may have to be borne if the HE occurs, and the inequality $0 < c_{min}$ indicates that, even in the best-case scenario, the HE will still result in some (potentially minor) losses. The latter assumption is particularly valid in the context of post-event mitigation analyses, as it acknowledges that even with successful mitigation, there may still be residual impacts or costs associated with the event.

Using $\Bold{c}$, $\Bold{w}$, and $\Bold{\overline{w}}$, we can now calculate the collective estimated risk $r$ and the collective tolerable risk $\overline{r}$, as shown earlier at \Eq{Eq_RiskDiscB}:
\begin{align}
&r=\Bold{w} \times \Bold{c}^T, ~~ \overline{r}= \Bold{\overline{w}} \times  \Bold{c}^T \label{Eq_RiskTolCol}
\end{align}
and the risk is considered tolerable if $r \leq \overline{r}$.

\begin{figure}[!h]  
\begin{center}
\includegraphics[scale=0.25]{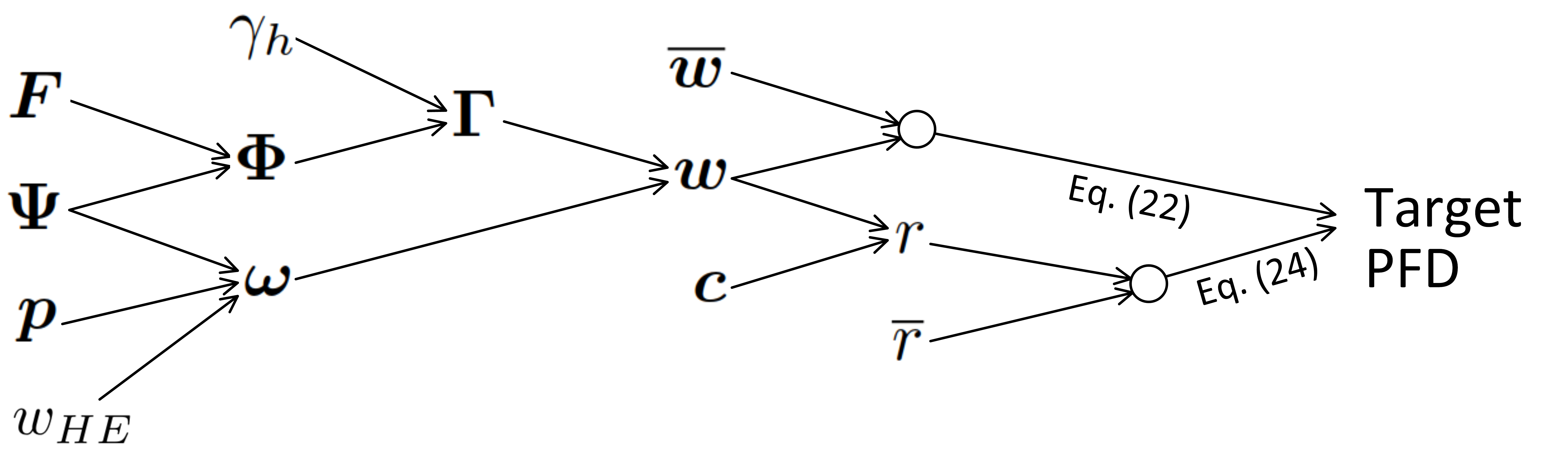}
\end{center}
\caption{Risk Estimation Process}
\label{Fig_Process}
\end{figure}

The risk estimation process we formulated here can be summarised as shown in \Fig{Fig_Process}. This figure shows the flow of calculation from left to right, with the final result being the target PFD. 

In a SIL allocation analysis, whether we use \Eq{Eq_RiskTolInd} or \Eq{Eq_RiskTolCol}, the target SIL will be determined by the gap between the tolerable and estimated levels of risk. Generally, our SIL allocation method comprises the following steps: 

\begin{enumerate}
\item Include the SF as an existing function in the risk estimation, using a conservative, realistic setting for its PFD; 
\item calculate the estimated risk and compare it with the tolerable level of risk, as in \Eq{Eq_RiskTolInd} or \Eq{Eq_RiskTolCol};
\item if the tolerance risk level is not met, reduce the elements of $\Bold{p}$ that corresponds the SF. Repeat as needed.
\item Use the final values in $\Bold{p}$ as the target PFDs for the SF subsystems.
\end{enumerate}

In the following section, we demonstrate the application of our method  through a realistic case study.

\section{Case Study}\label{Sec_CaseStudy}

The hazardous event we study here is a fire scenario in a 6km twin road tunnel. Suppose the frequency of occurrence of fire in tunnel is $w_{HE}=\V{0.7}$ per year, and the tunnel operation company uses the risk criteria given in \Table{Table_Cons}.\footnote{The figures presented in this case study have been adjusted to safeguard the confidentiality of project information. The provided data serves illustrative purposes while remaining realistic.} In this table, the first column indicates the severity segments and the last column their tolerable frequencies. From \Table{Table_Cons}, we have $n=5$ and $\overline{\Bold{w}}=\bigl[\V{0.001} ~~ \V{0.01} ~~ \V{0.1} ~~ \V{1} ~~ \V{10} \bigr]$.

\begin{table}[htbp]
\begin{center}
\begin{tabular}{|l|l|l|l|}
\hline \multicolumn{1}{|c|}{\TE{Conseq. Seg.}}&\multicolumn{1}{c|}{\TE{Safety}}&\multicolumn{1}{c|}{\TE{Financial}}&\multicolumn{1}{c|}{\TE{Tolerance Freq.}}\\ %Table top line
\specialrule{0.2em}{0.0em}{0.0em} % Title row separator
\TE{Catastrophic} & \TE{Fatalities} & \TE{$\geq \V{\$40M}$} & \TE{0.001 p.y.} \\ % Row 1
\TE{Major} & \TE{Permanent disabilities} & \TE{$\V{\$20} - \V{\$40M}$} & \TE{0.01 p.y.} \\ % Row 2
\TE{Moderate} & \TE{Hospitalisation} & \TE{$\V{\$5} - \V{\$20M}$} & \TE{0.1 p.y.} \\ % Row 3
\TE{Minor} & \TE{Medical treatments} & \TE{$\V{\$0.5M} - \V{\$5M}$} & \TE{1 p.y.} \\ % Row 4
\TE{Insignificant} & \TE{First aid} & \TE{$\leq \V{\$0.5M}$} & \TE{10 p.y.} \\ % Row 5
\hline %Table bottom line
\end{tabular}
\end{center}
\caption{Current Parameters Setting}
\label{Table_Cons}
\end{table}

The functions that are triggered in fire situations include the following:
\begin{itemize}
\item AFS: Automatic Fire Suppression
\item MFS: Manual Fire Suppression
\item ASE: Automatic Smoke Extraction
\item MSE: Manual Smoke Extraction
\item EE: Emergency Evacuation
\end{itemize}

Out of these functions, ASE is the candidate function for which we would like to determine a target PFD and SIL. 

The subsystems that perform the above mitigation functions are as follows:
\begin{itemize}
\item LHD: Line Heat Detector, detecting fire by sensing heat
\item FDP: Fire Detection Panel, interfacing between the LHD and other systems
\item IAD: Intelligent Accident Detector, including cameras and image processing algorithms, capable of alarming on fire incidents
\item PCS: Plant Control System
\item TOp: Tunnel Operator
\item OMS: Operation Management System, interfacing between operator and PCS
\item FSS: Fire Suppression System, including deluge valves
\item TVS: Tunnel Ventilation System, extracting smoke out of the tunnel
\item EMS: Egress Management System, including emergency lighting and signage 
\item TUs: Tunnel Users (i.e., drivers and passengers)
\end{itemize}
  
With $m=5$ functions and $l=10$ subsystems, the mapping matrix $\Bold{F}$ is defined as follows: 

\[
 \newcommand\col[1]{\parbox{1cm}{\centering\scriptsize#1}}
 \Bold{F} = 
 \begin{blockarray}{cccccc}
 AFS & MFS & ASE & MSE & ~EE & \\
 \begin{block}{[ccccc]l}
 1 & 0 & 1 & 0 & 0 & ~LHD \\
 1 & 1 & 1 & 0 & 0 & ~FDP \\
 0 & 1 & 0 & 0 & 0 & ~IAD \\
 0 & 1 & 1 & 1 & 1 & ~PCS \\
 0 & 1 & 0 & 1 & 1 & ~TOp \\
 0 & 1 & 0 & 1 & 1 & ~OMS \\
 1 & 1 & 0 & 0 & 0 & ~FSS \\
 0 & 0 & 1 & 1 & 0 & ~TVS \\
 0 & 0 & 0 & 0 & 1 & ~EMS \\
 0 & 0 & 0 & 0 & 1 & ~TUs \\
 \end{block}
 \end{blockarray}
\] 
As an example, the first column in $\Bold{F}$ shows that the Automatic Fire Suppression function relies on three subsystems: LHD as sensor, FDP as processing element, and FSS as the acting element. Accordingly, the failure of any of these three subsystems will result in the failure of AFS.

\begin{figure}[!h]  
\begin{center}
\includegraphics[scale=0.45]{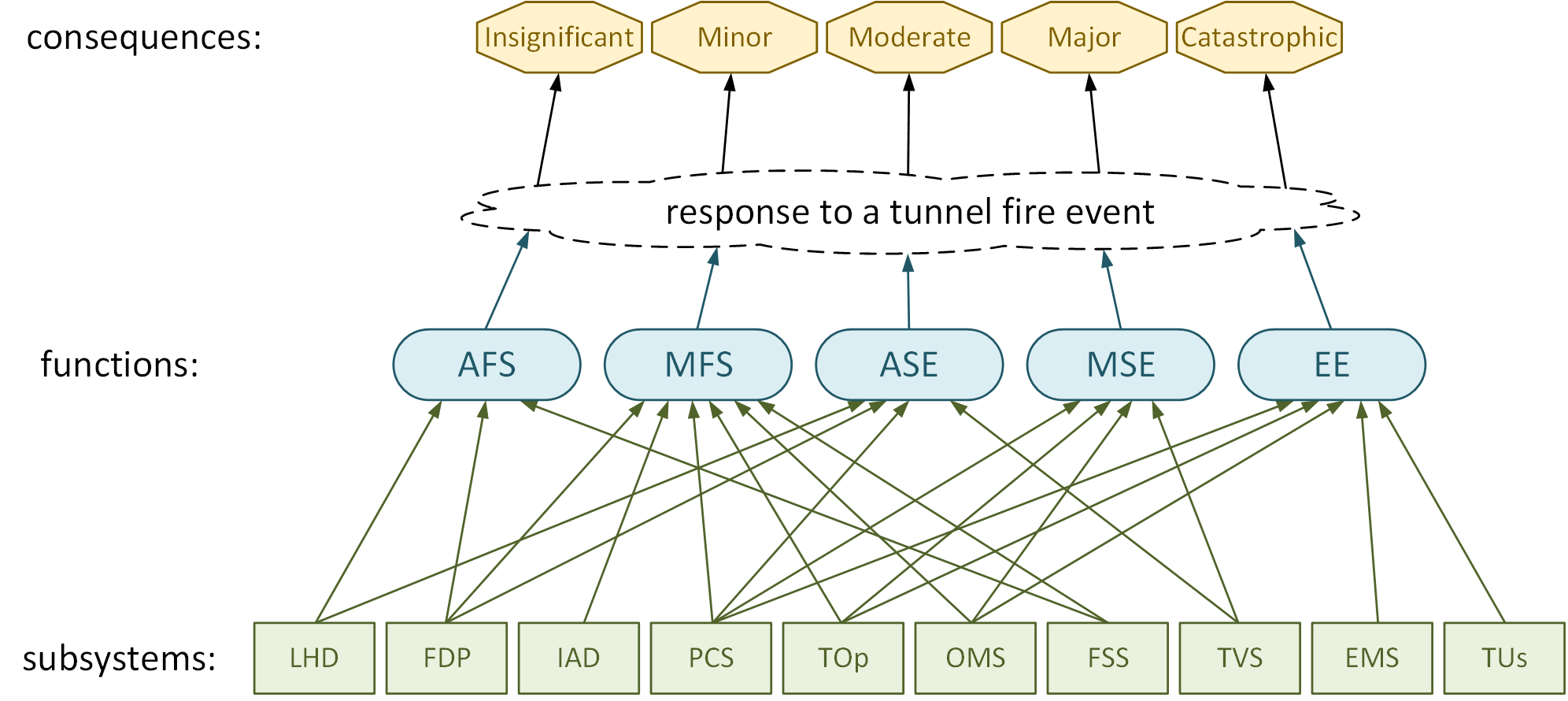}
\end{center}
\caption{Relationships Between Subsystems, Functions and Consequences}
\label{Fig_Flow_Corrected}
\end{figure}

The relationships between the subsystems and the mitigation functions are graphically depicted in \Fig{Fig_Flow_Corrected}, along with a simplified overview of the entire mitigation scenario: The failure/success of subsystems determines the failure/success of the functions. The functions are deployed in parallel in response to the fire event. Depending on the effectiveness of these functions, various consequences may arise to different extents.

Given $l=10$, the system state matrix $\Bold{\Psi}$ will be a 1024$\times$10 truth table that, in combination with $\Bold{F}$, generates a 1024$\times$5 function state matrix $\Bold{\Phi}$, as shown earlier in \Eq{eq_MatrixAND} and \Eq{eq_MatrixAND1}.

The next step will be to construct the state consequence matrix $\Bold{\Gamma}$, for which we define the logic functions $\gamma_h(i, \Bold{\Phi})$ as follows:
\begin{align}
\gamma_1(i, \Bold{\Phi})=&\neg \phi_{i3} \wedge \neg \phi_{i4} \wedge \neg \phi_{i5} \nonumber \\
\gamma_2(i, \Bold{\Phi})=&\neg \phi_{i1} \wedge \neg \phi_{i2} \wedge \neg \phi_{i5} \wedge \neg \gamma_1(i, \Bold{\Phi}) \nonumber \\
\gamma_3(i, \Bold{\Phi})=&\neg \phi_{i1} \wedge \neg \phi_{i2} \wedge \phi_{i5} \nonumber \\
\gamma_4(i, \Bold{\Phi})=&\neg \gamma_3(i, \Bold{\Phi}) \wedge \neg \gamma_2(i, \Bold{\Phi}) \wedge \neg \gamma_1(i, \Bold{\Phi}) \nonumber \\
\gamma_5(i, \Bold{\Phi})=&0 \nonumber
\end{align} 

Functions $\gamma_{1}$ to $\gamma_{5}$ represent different consequences, as follows:

\begin{itemize}
\item $\gamma_{1}$ represents the Catastrophic consequence scenario where both emergency evacuation and smoke extraction have failed. This scenario corresponds to the worst-case situation, where people are exposed to extensive smoke, leading to the potential asphyxiation and loss of lives.

\item $\gamma_{2}$ represents the Major consequence scenarios where smoke extraction is successful but people are still exposed to the fire. Such scenarios may result in burns and permanent disabilities for individuals.

\item The scenario $\gamma_{3}$ reflects the Moderate financial impact. In this scenario, emergency evacuation is successful, and people are moved to safety; however, the failure of fire suppression systems leads to equipment damage, resulting in financial losses.

\item For scenarios that do not fall into the above categories, we still consider Minor financial impacts due to operational interruption and potential tunnel closure. This is represented by $\gamma_{4}$.

\item Finally, $\gamma_{5}=0$ indicates that no fire incident is considered to have only insignificant consequences.
\end{itemize}
Using $\gamma_{1}$ to $\gamma_{5}$, matrix $\Bold{\Gamma}$ can be generated to indicate the classification of consequences in individual system states. 

We now need to calculate the individual state frequencies $\Bold{\omega}$, for which we need to know the PFD values of the subsystems. Our candidate SF is the ASE function, which is performed by subsystems LHD, FDP, PCS, and TVS. The target PFD for ASE is determined through a trial-error process where the $\Bold{p}$ parameters are optimised. We start with a realistic, conservative assignments to $\Bold{p}$ and calculate the risk. If the risk is found higher than the tolerable limit, we adjust those elements of $\Bold{p}$ that are related to ASE. The iteration process is terminated when the estimated risk is lower than the tolerance threshold. 

In setting up the PFD vector $\Bold{p}$, we can use four independent variables representing the PFD of the LHD, FDP, PCS, and TVS subsystems, which constitute the SF subsystems. Alternatively, a simpler approach is to apportion these values to a single parameter $p$. This reduces the degrees of freedom and aids in setting the overall target PFD that we are interested in. 

This apportionment can be based on industry experience and the proportion of failures observed in similar installation environments. In this case study, we know from experience that LHD, FDP, PCS, and TVS subsystems may have a rough proportion of $\V{25\%}$, $\V{20\%}$, $\V{20\%}$, and $\V{35\%}$ of the overall PFD, respectively. Therefore, we define $\Bold{p}$ as follows:
\begin{align}
\Bold{p}=\bigl[\V{0.25}p ~~ \V{0.2}p ~~ \V{0.05} ~~ \V{0.2}p ~~ &\V{0.1} ~~ \V{7E-4} ~~ \V{0.04}  ~~ \V{0.35}p ~~ \V{0.02} ~~ \V{0.2} \bigr] \label{Eq_PFDs}
\end{align}
where $p$ indicates the target PFD of the SF. The PFD values of the other subsystems are set based on prior experience and by taking a realistic, conservative approach.

Let the starting point be $p=\V{0.1}$ which is the boundary value for SIL1. This initial allocation can make it clear in the first iteration whether a SIL ranking will be required at all. If so, further adjustments can be made to optimise the value of $p$. With $p=\V{0.1}$, we will have:
\begin{align}
\Bold{p}=\bigl[\V{0.025} ~~ \V{0.02} ~~ \V{0.05} ~~ \V{0.02} ~~ &\V{0.1} ~~ \V{7E-04} ~~ \V{0.04}  ~~ \V{0.035} ~~ \V{0.02} ~~ \V{0.2} \bigr] \nonumber
\end{align}

Given $w_{HE}$, $\Bold{p}$ and $\Bold{\Psi}$, we can use \Eq{Eq_SecenFreq} to calculate the frequency of individual states ($\Bold{\omega}$). Then, by using $\Bold{\omega}$ and $\Bold{\Gamma}$, we can calculate $\Bold{w}$, the consequence frequencies:
\begin{align}
\Bold{w}=\bigl[\V{2.4E-02} ~~ \V{1.03E-02} ~~ \V{2.92E-02} ~~ \V{6.36E-01} ~~ \V{0} \bigr] \nonumber
\end{align}

Upon comparing $\Bold{w}$ with $\overline{\Bold{w}}$, it becomes evident that, with a target PFD of $\V{0.1}$ for the ASE function, the estimated frequency of consequences in the Catastrophic and Major categories exceeds the tolerance thresholds. Our objective now is to determine the target PFD that ensures $w_k \leq \overline{w}_k$.

After a couple of  iterations, we have arrived at a target PFD of $p=\V{4E-03}$, which leads to the following consequence frequencies:
\begin{align}
\Bold{w}=\bigl[\V{9.75E-04} ~~ \V{8.34E-03} ~~ \V{2.01E-02} ~~ \V{6.71E-01} ~~ \V{0} \bigr] \nonumber
\end{align}
in which the consequence frequency in each individual segment is below the tolerance threshold, indicating the end of our SIL allocation. Our target PFD is $p=\V{4E-03}$, which falls within the range of SIL2.

\section{Literature Review}\label{Sec_LitReview}

There is no shortage of resources on SIL determination. The topic has been around for long enough to have industry standards \cite{Ref_183,Ref_190,Ref_347,Ref_348,Ref_349,Ref_354}. Numerous research works have been conducted to cover various aspects, including setting out the concepts \cite{Ref_359}, collecting surveys \cite{Ref_350}, addressing the impact of cost \cite{Ref_345,Ref_370}, and addressing uncertainty \cite{Ref_357}.\footnote{The references cited here are only provided as samples, not suggesting that they are the only available works.}  However, SIL determination for \emph{mitigation} functions has rarely been addressed. This section reviews the works most relevant to this topic. 

ISA's technical report TR84.00.07-2018 \cite{Ref_368} offers a comprehensive guideline on a specific type of mitigation functions - Fire and Gas Systems (FGSs). This guideline considers the FGS as a typical Safety Related System \cite{Ref_186}, comprising sensors, logic solver, and final elements. However, it defines the overall ``performance'' of the FGS as a combination of the PFD, sensor coverage, and the effectiveness of the final elements.

The TR84.00.07 clarifies that LOPA is not the appropriate method for analysing the performance of an FGS for two main reasons: LOPA is suitable for preventive functions where the outcome is either complete success or complete failure, and LOPA necessitates independence between the protection layers and the initiating events, which may not always be attainable in fire and gas scenarios. Consequently, the TR84.00.07 employs ETA to estimate the frequency of consequence and to express the overall effectiveness of the FGS.

In comparison, our approach does align with some underlying concepts of TR84.00.07, but it diverges from this guideline in some aspects: TR84.00.07 is specific to FGS, while our method is generic; TR84.00.07 covers all aspects of an FGS, including those not related to SIL and PFD, whereas our generic method focuses solely on determining target PFDs; and TR84.00.07 employs ETA, requiring independent protection mechanisms, whereas our method allows for shared subsystems. 

Another body of work, including \cite{Ref_362,Ref_363,Ref_364,Ref_365}, has focused on quantitative risk assessment in escalation and domino scenarios. These scenarios typically commence with the occurrence of one hazardous event (e.g., a fire) and are primarily concerned with minimizing the likelihood of escalation to a second hazardous event.

Similar to TR84.00.07, the aforementioned research works utilise ``performance'' indicators to assess the efficacy of safety barriers. These indicators encompass both the PFD and the effectiveness of the barrier. The distinction between domino scenarios and those addressed by TR84.00.07 lies in the fact that in the former, the mitigation function averts a secondary event if successful, whereas in the latter, the mitigation function merely mitigates the severity of the consequence.

These research efforts distinguish between the PFD as an indicator of unavailability and ``effectiveness'' as a probabilistic measure for the likelihood of an available mitigation function successfully preventing the secondary incident. To illustrate this distinction, consider a firefighting system used to suppress a fire in a first fuel tank to prevent it from spreading to a second tank. Even if all firefighting equipment and personnel are available to respond, the effectiveness of the measure still depends on factors such as the response delay time to the initial incident. Such external factors impact the effectiveness of the barrier and should be considered independently of the system's availability.

These studies, too, utilise ETA to estimate the frequency of potential outcomes, typically limited to three segments: prevented escalation, mitigated escalation, and full escalation. However, the primary focus of these research works is on specific applications (e.g., fire or flood), rather than SIL determination. In fact, they typically begin their analyses through techniques such as FTA \cite{Ref_362} or conservative approximations \cite{Ref_363} to estimate the PFD first, which is then used to calculate and evaluate the overall performance of mitigation.

Other studies have also employed ETA and frequency of consequence to evaluate post-incident risk reduction. For instance, \cite{Ref_366} presents an FGS case study in which two ETAs, with and without the Safety Related System, are utilised to estimate the frequency of consequence and subsequently assess the risk. The outputs of these analyses are then compared to determine the target PFD for the SIL-rated FGS function.

While this approach may be effective in small-scale applications, constructing an ETA for larger applications presents challenges. These challenges stem not only from the potential size of the tree becoming unmanageable but also from the potential dependencies between subsystems, which can compromise the validity of the ETA. The method presented here addresses both of these challenges and offers an application-independent approach. 

As another example, \cite{Ref_367} presents a case study from a sub-sea application where ETA is utilised to calculate the frequency of various outcomes. In this study, two independent Safety Functions rated SIL2 are employed, both of which may be activated in a hazardous scenario. However, the primary focus of this study is on comparing PFD and PFH in meeting a target SIL and deciding on further risk reduction measures, rather than determining the target SIL itself. 

One challenge addressed in this research is that the demand rate is close to the boundary of 1 per year, which distinguishes between Low Demand and High Demand mode SFs, for which PFD or PFH may be more suitable. The paper utilises the case study and the PFD and PFH calculations to compare the feasibility of the two measures and to evaluate the control measures that can be implemented to practically meet the targets.

Using frequency as the core measure in risk analyses, instead of probability, offers another advantage with respect to SF operation modes: The PFD and PFH use different formulations for calculation, but both depend on the same frequency of failure. Calculating risk in frequency terms offers a unified analysis approach, independent of the SF mode \cite{Ref_374,Ref_375}. This paper takes the same approach to use frequency as the key element; however, our work addresses mitigation functions, rather than the issue of different SF modes.

Finally, a body of work has focused on optimisation methods used in the context of risk assessment and risk management \cite{Ref_345,Ref_370,Ref_371,Ref_372,Ref_373}. In general, the main objective of these studies is to find a set of parameters that can minimise a target criterion (e.g., a combination of risk and cost). In the context of this paper, optimisation can be related to the problem of apportioning the target PFD to multiple subsystems such that a risk limitation is met and the cost is minimised. However, the focus in the current paper is on the underlying concepts and formulation. Mathematical PFD appointment methods can be addressed in future works.

\section{Discussion}\label{Sec_Discussion}

One key difference between preventive and mitigation SFs lies in how the failure state of a function is characterized. A preventive SF is deemed successful if it prevents an event and unsuccessful if it fails to do so. In contrast, the success and failure of mitigation SFs can be measured on a continuous scale based on the degree to which they reduce the consequences.

The challenge here is that allocating a target PFD is only relevant in scenarios where the failure is expressed in binary form, as a single probability value (PFD) can only distinguish between two disjoint outcomes of the same event.

Our goal in this paper was not to debate the conceptual suitability of SIL for mitigation scenarios but to enhance the methodological aspect of it. When applying our method, the degree of failure must be partitioned into two disjoint segments. For example, consider a ventilation damper consisting of multiple independent louvres, for which we define 60\% as the threshold between successful and failed opening. In this case, the PFD will represent the probability that more than 40\% of the louvres fail to open when there is a demand.

In the remainder of this section, we briefly discuss other aspects of SIL determination for mitigation functions.

\subsection{Methodological Necessity}\label{Sec_Need}

Some of the current methods used for SIL allocation - RG, LOPA, FTA, and ETA - were discussed in Section \ref{Sec_Background}. Let us summarise why a new method is required for allocating target PFD/SIL to mitigation functions.

Risk Graph is a quick and easy qualitative approach but it has some downsides. It is subjective and inconsistent, as the assessment relies on expert judgement, leading to variability in results between different assessors. It also lacks mathematical rigour, making it difficult to justify decisions in highly regulated industries. Additionally, it forces discrete categories, which may not accurately capture real-world risk distributions, and it may lead to overestimation or underestimation of SIL. 

More specifically related to mitigation functions, RG is unsuitable because mitigation scenarios are often complex, involving multiple interacting hazards mitigated by several functions in parallel. RG does not explicitly account for individual protection layers, making it difficult to separate between the risk reduction required by one function among several.

LOPA is a semi-quantitative method that accounts for individual functions. However, it is primarily designed for preventive functions, as it is limited to single-consequence scenarios. Additionally, LOPA requires independence between all protection layers, which is difficult to achieve in complex mitigation scenarios, such as the case study presented earlier.

The quantitative approach of FTA can overcome the independence condition limitation in LOPA. However, FTA, too, is primarily suitable for prevention scenarios, where multiple causes of the HE interact with multiple (potentially dependent) functions, resulting in a single HE. When used for mitigation scenarios, the tree should be adjusted such that: all basic events are expressed in terms of fixed probability events; multiple top events are modelled to represent multiple consequence scenarios - which is not typical in a standard FTAs; and the HE is modelled as a frequency event AND'd with the final top events. Still, the results may be underestimated, compared to the simple product formula used in LOPA and ETA - because of the the way in which FTA tools often calculate top event frequencies \cite{Ref_168}.

FTA is primarily designed to model failure events of components, and tools are often adjusted based on that approach. Special care should be taken when FTA tools are used solely for probability or frequency calculations. Furthermore, using FTA often requires expensive tools and specialised skills.

Among the classic methods, ETA is the most relevant for mitigation scenarios. It begins with a single HE and considers multiple consequences, representing the left-hand side of the Bowtie model. However, there are two limitations that may make it inefficient for SIL allocation in complex scenarios. First, ETA, too, requires independence between the protection layers, which our case study clearly showed is not always feasible in practical scenarios. Second, ETA is not primarily a SIL allocation method. If used, additional steps should be taken to estimate the risk based on calculated frequencies and compare it to the risk criteria for target PFD estimation. Furthermore, ETA also requires specialised tools and skills, and manually creating Event Trees for complex scenarios can be time-consuming and prone to errors.

Where the SF being studied is a mitigation function entangled with other functions to mitigate complex hazard scenarios, qualitative methods (e.g., RG) and those quantitative methods that rely on the independence condition (e.g., LOPA and ETA) are not suitable. Using FTA comes with its own downsides, as outlined above. Our method borrows conceptual ideas from these commonly used methods and offers a solution that is both precise and easy to use. The method is not limited by the independence condition, as it incorporates an additional layer of subsystems. It does not require specialised tools or skills and can be implemented using generic spreadsheet tools. It is more precise, as it considers all system states without the need to build a complicated tree, and is suitable for complex mitigation scenarios, regardless of the number of functions, subsystems, or states.

\subsection{PFH and High Demand Mode SFs}

As defined by the standard \cite{Ref_186}, SFs can operate in either Demand mode or Continuous mode. Demand mode SFs can be further classified into two groups: Low Demand, where the demand rate is no greater than one per year, and High Demand, where the demand rate exceeds one per year. To distinguish between these categories, the standard requires the use of PFD for Low Demand SFs and PFH for Continuous and High Demand SFs \cite{Ref_183}. Although failure frequency (PFH) is not a direct reflection of the reliability of a system that is inherently of Demand mode type, it is still preferable to use PFH in such scenarios because the other option (PFD) can lead to over-engineering of the SF \cite{Ref_361}.

Mitigation SFs inherently fall under the Demand mode type, as they only respond to a demand triggered by a hazardous event. Consequently, the possibility of Continuous mode for these SFs can be ruled out. However, High Demand mitigation SFs are still feasible (e.g., when the fire frequency is high) for which PFH needs to be determined as target. 

When calculating the actual availability measure for an existing system configuration, PFD and PFH are not directly convertible to each other. However, when setting \emph{targets}, one has the flexibility to decide on a target PFH based on a desired target PFD, and vice versa.

Consider a single component with a ``mission time'' of $\tau$. Where the component is proof tested, $\tau$ is the proof test interval. Otherwise, the mission time will be the period during which undetected faults of component remain undetected. Let $\lambda_{DU}$ be the dangerous failure rate of the component. It can be shown that for that single component \cite{Ref_188}, \cite{Ref_360}:
\begin{align}\label{Eq_PFHPFD}
PFH = &\lambda_{DU} \\
PFD = &(\lambda_{DU} \cdot \tau)/2, ~~ for ~ \lambda_{DU}\cdot  \tau \ll 1
\end{align}
 
For the purpose of setting targets, and where $PFH \cdot \tau \ll 1$, the following can be used to relate target $PFD$ and $PFH$:
\begin{align}\label{Eq_PFHPFD}
PFH = (2 PFD) / \tau 
\end{align}

Therefore, when interested in setting target PFH, instead of PFD, the analyst can still use the new method introduced in \Sec{Sec_PFDFormul} to calculate target PFD and then use \Eq{Eq_PFHPFD} to calculate the corresponding target PFH. 

For the case study covered earlier, the target PFD was calculated as $\V{2.1E-03}$. Suppose we want to calculate the target PFH in this scenario, even though the demand rate is less than one per year. Assume that the ASE function is proof tested once a year, meaning that $\tau=\V{8760}$ hours. Hence, the target PFH in this case will be $\V{4.8E-07}$, which is also in the PFH range for SIL2 as given in the standard \cite{Ref_183}.

\subsection{Coverage and Effectiveness}

This paper was focused on formulating the probabilistic aspect of \emph{failure} in mitigation functions. Another crucial aspect of a mitigation function is its relative efficacy in detecting hazards and executing the SF action, even when it is in the success state.

ISA's technical report TR84.00.07 \cite{Ref_368} addresses sensor coverage and final element effectiveness for FGSs. The method employed in this TR incorporates these parameters as two deterministic decimal values. Alongside PFD, these values are utilised to derive a ``performance'' factor, which is then employed to assess the overall effectiveness of the FGS. In other studies, such as \cite{Ref_362,Ref_363,Ref_364,Ref_365}, the effectiveness of a post-accident preventive function is integrated as a probabilistic factor, also distinct from PFD.

As sensor coverage and final element effectiveness are specific to application, they were not included in the general formulation presented in this paper. Nevertheless, if necessary, the deterministic approach outlined in TR84.00.07 or the probabilistic approach referenced in \cite{Ref_362,Ref_363,Ref_364,Ref_365} may be utilised to adapt our methodology accordingly.

\section{Conclusion}\label{Sec_Conclusion}

This paper delineated the inadequacies of current methodologies in allocating target PFD for mitigation Safety Functions, formulating a novel approach that could overcome the obstacles. The method was also substantiated with a practical case study centred on road tunnel fire scenarios.

The principal challenges addressed in this paper stem from the disparities between mitigation and preventive functions. These include the differences in the way these functions affect the risk and the intricacies of subsystem dependencies.

Our methodology tackles complex analyses involving multiple interdependent functions. Yet, it utilizes straightforward formulation suitable for implementation through commonly used spreadsheet tools. This accessibility broadens the user base, facilitating the practical application of the SIL allocation methods across diverse Safety Related Systems. The proposed method enhances existing methodologies while remaining fully compatible with current SIL concepts, ensuring seamless integration into established practices. 

In addition, exploring the use of optimisation techniques to apportion the optimal target probability parameters while minimising implementation costs can be considered a topic for further research.

\bibliographystyle{ieeetr} 
\balance
\bibliography{References}

\end{document}